\documentclass[a4paper,pra,twocolumn,showpacs,superscriptaddress]{revtex4-1}
\usepackage{epsfig}
\usepackage{amsbsy,latexsym}
\usepackage{amsmath}
\usepackage{amsthm}
\usepackage{amssymb, mathrsfs}
\usepackage[mathscr]{eucal}
\usepackage{color}
\usepackage[normalem]{ulem}
\newcommand{\tcb}[1]{#1}

\def\isom{\kappa}

\def\ket#1{|#1\>}
\def\bra#1{\<#1|}

\newcommand{\hilb}[1]{\mathcal{#1}}

\def\<{\langle}
\def\>{\rangle}

\newcommand{\tr}[1]{\mathrm{tr}[#1]}

\begin{document}

\title{Private quantum channels for multi-photon pulses and unitary k-designs}

\author{Jan Bouda}
\affiliation{Faculty of Informatics, Masaryk University, Botanick\'{a} 68a,
             602\,00 Brno, Czech Republic}
\email{3717@mail.muni.cz}

\author{Michal Sedl\'ak}
\affiliation{Centre of Excellence IT4Innovations, Faculty of Information Technology, Brno University of Technology, Bo\v zet\v echova 2/1, 612 00 Brno, Czech Republic }

\author{Mario Ziman}
\affiliation{Faculty of Informatics, Masaryk University, Botanick\'{a} 68a,
             602\,00 Brno, Czech Republic}
\affiliation{Research Center for Quantum Information, Institute of Physics,~Slovak Academy of Sciences,~D\'ubravsk\'a cesta 9,~84511 Bratislava,~Slovakia}

\date{\today}

\begin{abstract}
  We address the question of existence of private quantum channel for qubits encoded in polarization degrees of freedom of a photon, that remains secure even if multi-photon (instead of single-photon) pulse is emitted. We show that random unitary channel distributed according to  SU(2) Haar measure has this property. Further we analyze the qubit unitary k-designs. We show they ensure security if the photons' parity of the source is guaranteed. Otherwise, the qubit unitary k-designs do not guarantee perfect security.
 \end{abstract}

\maketitle

\noindent
\section{Introduction}
Symmetric encryption of messages is a fundamental and well studied problem in classical cryptography. The goal is to encode message (plaintext) into a ciphertext in such a way that legitimate parties (usually called Alice and Bob) can perfectly recover the original message, while their adversary (traditionally called Eve) gains no (additional) knowledge about the plaintext from the ciphertext.

\begin{figure}[t]
\centerline{\includegraphics[width=0.8\linewidth]{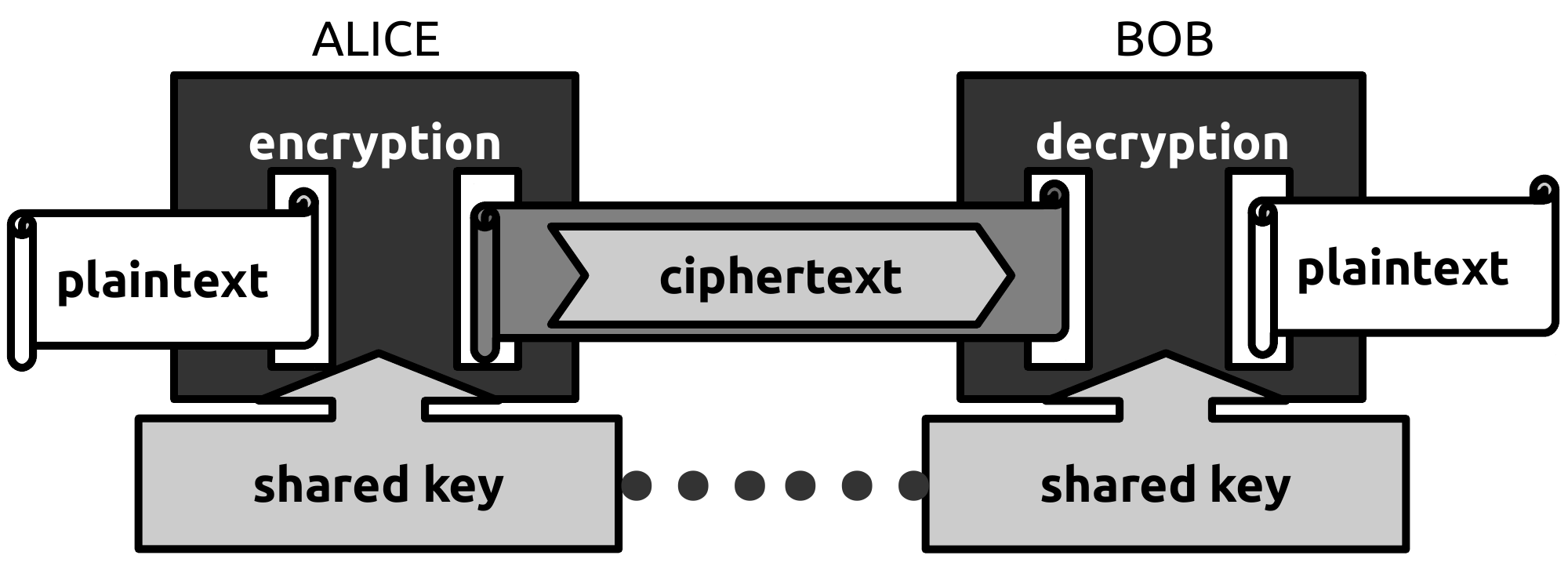}}
\caption{Symmetric encryption.}
\label{fig:1timepad}
\end{figure}

In order to implement encryption, Alice and Bob must have some advantage when compared to Eve. It is impossible to achieve secure encryption in a perfectly symmetric situation. The standard advantage Alice and Bob have is a shared secret bit string (``key"),  see Figure \ref{fig:1timepad}.

One-time pad (also known as Vernam cipher) is an encryption system that achieves information-theoretical security, namely the security is not based on any computational assumptions. The encryption and decryption operation are the same, bitwise application of XOR (control-not) operation controlled by the key, and applied to plaintext or ciphertext, respectively. The key is a uniformly random bit string, shared between Alice and Bob, and unknown to Eve. Each bit of the key can be used only once, i.e. to encrypt only single bit of plaintext. This is where the name one-time pad comes from. In fact, this property is necessary for any information-theoretically secure encryption scheme \cite{shannon1949}.

The quantum generalization of one-time pad give raise to various quantum communication primitives depending on which part is ``quantized" (plaintext, or ciphertext, or or the shared key). Quantum teleportation \cite{1993bennett} and superdense coding \cite{1992bennett} protocols are the most prominent examples. In particular, the teleportation is understood as secure transfer of quantum plaintext by transfering classical ciphertext and exploiting quantum key. Accordingly, the goal of superdense coding is to employ quantum key to securely transfer classical plaintext by means of quantum ciphertext.

In this paper we address the question of secure transfer of quantum plaintext by communicating quantum ciphertext encrypted by classical key. This situation was studied under the name private quantum channel \cite{2000ambainis}, resp. quantum one-time pad \cite{2003boykin}.

For private quantum channels the security is guaranteed by a shared classical secret key. It was shown in \cite{2000ambainis,2003boykin} that two bits of a secret key are necessary and sufficient to encrypt one qubit, i.e. one two level quantum system. In particular, based on the shared value of two bits (i.e. values $0,1,2,3$), Alice and Bob apply the encryption and decryption operations by implementing the corresponding Pauli unitary gates $\sigma_0=I,\sigma_1=\sigma_x,\sigma_2=\sigma_y,\sigma_3=\sigma_z$. Alice applies this unitary onto a qubit (quantum plaintext), which should be encrypted. This yields a quantum ciphertext, which is transmitted via a quantum channel to Bob. The decoding is simply done by inverting the unitary that was applied by Alice, which is possible due to existence of the pre-shared secret key.
The full security of the quantum one time pad protocol is guaranteed if the secret key is uniformly random, the key is used only once and if the implemented operations are really unitary, thus, the following identity holds
$$
\varrho\mapsto{\cal E}(\varrho)=\frac{1}{4}(\varrho+\sigma_x\varrho\sigma_x
+\sigma_y\varrho\sigma_y+\sigma_z\varrho\sigma_z)=\frac{1}{2}I\,.
$$
This means that from the point of view of an eavesdropper any qubit of a plaintext is mapped into a completely mixed state $\rho^{(0)}\equiv\frac{1}{2}I$.

Assuming the dimension of the plaintext equals dimension of the ciphertext, the only invertible operations are unitaries. In order to encrypt the plaintext, we choose randomly (according to the key) a unitary operation, encrypt the plaintext, transmit it, and decrypt it by the inverse operation. The essence of security is that without knowledge of the key, the ciphertext carries no information about the plaintext (for details see definition in \tcb{\cite{2000ambainis} and Appendix C}), i.e. the average state $\rho^{(0)}$ transmitted is independent of the plaintext $\rho$:
\begin{equation}
\label{eq:PQCdef}
    \exists \rho^{(0)}\ s.t.\ \forall\rho\ \sum_i p_i U_i\rho U_i^\dagger=\rho^{(0)}.
\end{equation}
It is known that $\rho^{(0)}=\frac{1}{2}I$ for any reasonable set of plaintexts, namely if $\frac{1}{2}I$ is in the convex span of plaintexts \cite{2007bouda}.

This gives us that the average encryption channel is the full Haar-random unitary channel sending all states to the completely mixed state. \tcb{However,} we need a finite (as small as possible) set of unitaries,  so this encryption scheme is equivalent to unitary $1$-design \cite{2007gross}, i.e.
$$
\sum_j q_j U_j\varrho U_j^\dagger = \int_{SU(2)}dU U\varrho U^\dagger\,,
$$
where $dU$ is the (group invariant) Haar measure over unitary channels. Let us stress that for
encryption of $N$ qubits the length of the key reads $l(\vec{q})=-N\sum_j q_j\log_2 q_j$. It achieves
its minimal value if and only if unitary channels $U_j$  are orthogonal ($\tr{U_j^\dagger U_k}=0$ for $j\neq k$)
and $q_j=1/4$ for $j=1,2,3,4$, i.e. optimal private quantum channel consists
of four (uniformly distributed) encryption/decryption operations.

The basic framework of private quantum channel can be generalized in various different ways. The  encryption of restricted sets of plaintexts was investigated in \cite{2006nayak,2007bouda}, the non-malleable encryption was addressed in \cite{2009ambainis} and the encryption protocols for quantum continuous variables
were designed and reported in \cite{2005bradler,2015jeong}.

In practice, when we try to implement quantum one time pad, some of the requirements may be hard to meet, hence, the security of the ideal case is challenged. For example, transmission of light seems to be nowadays the best option for sending information between distant parties. If qubits are encoded into single photons then unitary transformations can be performed quite reliably, but the preparation and the measurement of single photon are often inefficient and noisy. Our goal in this paper is to study and partially resolve security issues that may arise from quantum state preparation. More precisely, we consider the situation, when qubits are stored in the polarization of light and multiple photons (instead of only one) are generated at the plaintext source. As we shall see this effect needs to be taken into account when evaluating the security of the private quantum channel implementation.

\tcb{The situation we are addressing in our paper is that when the experiment is set up to apply a unitary $1$-design to a single photon state, and instead multi-photon state is generated, the resulting operation performed on the multiphoton state does not fulfill the definition \eqref{eq:PQCdef}. Namely, the average output state is not independent of the input, what makes the whole setup insecure. This is analogous to the QKD scenario, when QKD protocol is secure for single photon pulses, but becomes insecure as soon as multi-photon pulse is emitted.}

The paper is organized as follows: in the following Section II we will introduce the formalism 
for quantum states of light, in Section III we show that Haar distributed unitaries form a valid private quantum channel that remains secure even for multi-photon pulses. Section IV analyzes the security of unitary $k$-designs for the private encryption given the maximal number of photons of the multiphoton source is limited by the energy constraint to $k$. In particular, we show that unitary $k$-design based private quantum channels remain secure for multi-photon pulses, provided certain restrictions are met. The conclusions and discussions of the results are presented in Section V.

\section{Multiphoton pulses of polarized light}

Since free electric (and magnetic) field oscillates in the plane perpendicular to the direction of the propagating light wave, it has two 
evolving components, which behave as two harmonic oscillators. If we denote the two components as $x$ and $y$ according to the two directions perpendicular to the propagation direction, say $z$, then we may introduce creation and annihilation operators $a^\dagger_x, a^\dagger_y, a_x, a_y$.
In the free space or in the isotropic optical fibers the two oscillators are independent and simultaneous eigenstates of the number operators $n_x=a^\dagger_x a_x$, $n_y=a^\dagger_y a_y$ exist \cite{bookmandel}.
Let us denote the related Hilbert space by $\hilb{H}$ and an arbitrary normalized state in it can be expressed as
\begin{align}
\ket{\Psi}=\sum_{m,n=0}^\infty c_{mn}\ket{m,n},
\end{align}
where $c_{mn}\in \mathbb{C}$, $\sum_{m,n} |c_{mn}|^2=1$ and $\ket{m,n}$ is an eigenstate of both $n_x$ and $n_y$ with eigenvalues $m,n$, respectively. Single photon polarization states are states sharing just a single excitation between the two oscillators (also called modes).
The state with no excitations is called vacuum and we denote it as $\ket{\emptyset}\equiv \ket{0,0}$.
If the directions $x$, $y$ are referred to as \emph{horizontal}, and \emph{vertical}, respectively, then an arbitrary single photon pure state can be seen as a superposition of a horizontal $\ket{H}\equiv \ket{1,0}$ and a vertical $\ket{V}\equiv \ket{0,1}$ state
\begin{align}
\label{eq:psi1}
\ket{\psi}=\alpha \ket{H} + \beta \ket{V} =(\alpha a^\dagger_x + \beta a^\dagger_y) \ket{\emptyset}
\end{align}
If $n$ photons with the same polarization are generated, then the state is described as
\begin{align}
\label{eq:psin}
\ket{\Psi}=\frac{1}{\sqrt{n!}}(\alpha a^\dagger_x + \beta a^\dagger_y)^n \ket{\emptyset}=\sum_{k=0}^n \alpha^k \beta^{n-k} \sqrt{\binom{n}{k}} \ket{k,n-k}
\end{align}
The manifold of $n$-photon states belongs to the $n+1$ dimensional subspace $\hilb{H}_n$ spanned by vectors $\{\ket{k,n-k}\}_{k=0}^{n}$.

Consider now system of $n$ qubits, each prepared in a state $\ket{\varphi}=\alpha \ket{0} + \beta \ket{1}$. Then the overall state can be written as
\begin{align}
\label{eq:psion}
\ket{\varphi}^{\otimes n}=(\alpha \ket{0} + \beta \ket{1})^{\otimes n}=\sum_{k=0}^n \alpha^k \beta^{n-k} \sqrt{\binom{n}{k}} \ket{s_k},
\end{align}
where $\ket{s_k}$ are the totally symmetric states of $n$ qubits with $k$ zeros (and $n-k$ ones).
This suggests us a simple bijective isometry between $n$-photon subspace and the totally symmetric subspace of $n$ qubits, which is defined by mapping states $\ket{k,n-k}$ onto $\ket{s_k}$ $\forall k$ or vice versa. 
Let us define $\hilb{K}^+_{n}$ as the totally symmetric subspace of $n$ qubits and $\hilb{K}^+_{0}$ as some one dimensional Hilbert space. The above isometry describes an isomorphism between $\hilb{H}_n$ and $\hilb{K}^+_{n}$. Consequently, we obtain an isomorphism $\isom$ between $\hilb{H}$ and
\begin{align}
\hilb{K}\equiv \oplus_{n=0}^\infty \hilb{K}^+_{n},
\end{align}
since $\hilb{H}= \oplus_{n=0}^\infty \hilb{H}_{n}$.

Thus, a polarization of an electro-magnetic wave of a fixed frequency in a single spatial mode or in an optical fiber can be equivalently described in Hilbert space $\hilb{K}$, which is roughly speaking a direct sum of totally symmetric subspaces of different number of qubits. For notational convenience we follow the following conventions. We express pure states $\ket{\Phi}$ from $\hilb{K}$ as
\begin{align}
\label{eq:Psiarb}
\ket{\Phi}= \sum_{n=0}^\infty c_n \ket{\varphi_n},
\end{align}
where $c_{n}\in \mathbb{C}$, $\sum_{n} |c_{n}|^2=1$, $\ket{\varphi_n}\in \hilb{K}^+_n$ and we note that $\langle \varphi_n \ket{\varphi_m}=\delta_{nm}$ by definition of the inner product on the direct sum Hilbert space.

For any polarization state $\ket{\psi}$ from Eq. (\ref{eq:psi1}) the $n$-photon state with same polarization (given by Eq.(\ref{eq:psin})) is mapped by $\isom$ into $\ket{\varphi}^{\otimes n}\in\hilb{K}^+_{n}\subset \hilb{K}$ (see Eq. (\ref{eq:psion})). It is easy to see that any linear optical transformation
of the two polarization modes
\begin{align}
U \leftrightarrow \left(
    \begin{array}{c}
    b^+_x \\
    b^+_y
    \end{array}
\right)
=
\left(
    \begin{array}{cc}
    U_{xx}& U_{xy} \\
    U_{yx}& U_{yy}
    \end{array}
\right)
\left(
    \begin{array}{c}
    a^+_x \\
    a^+_y
    \end{array}
\right),
\end{align}
which induces a unitary transformation $U$ of the single photon polarization $U\ket{\psi}=\ket{\psi'}$, corresponds in isomorphism $\isom$ to a transformation
$U^{\otimes n }\ket{\varphi}^{\otimes n }=\ket{\varphi'}^{\otimes n }$ in $\hilb{K}^+_{n}$, when applied to $n$ photons of the same polarization.
Thus, any change of polarization $U$, induced typically by quarter (QWP) and half wave plates (HWP), is represented in our isomorphism $\isom$ by action of unitary operator $L_n(U) \equiv U^{\otimes n}$ in subspace $\hilb{K}^+_{n}$ for every $n$. This corresponds to an overall unitary transformation $L(U)=\oplus_{n=0}^\infty L_n(U)=I\oplus L_1(U)\oplus L_2(U)\oplus \ldots$ acting in $\hilb{K}$. The mapping $U\in U(2) \mapsto L_n(U)\in\hilb{L}(\hilb{K}^+_{n})$ is an irreducible representation of the unitary group $U(2)$ of the order $n/2$, which is often referred to as a spin $n/2$ in physics.
In effect, the invertible encryption operation $U$ induced by QWPs and HWPs results for the multi photon pulse in the action of the direct sum representation $L(U)$ in Hilbert space $\hilb{K}$.

\section{Haar distributed unitaries and multiphoton source}
\label{haarandsource}
\tcb{Recall that the source is not ideal and prepares multi-photon state}
$$\ket{\Psi}=\sum_{n=0}^\infty c_n \frac{1}{\sqrt{n!}}(\alpha a^\dagger_x + \beta a^\dagger_y)^n \ket{\emptyset}\,.$$

\tcb{As explained before, plaintext is encoded into the polarization (amplitudes $\alpha, \beta$), and numbers $c_k$ do not contain any information on the polarization.}

In the isomorphism $\isom$, which we will use for all further descriptions, this state corresponds to a state in Eq.~(\ref{eq:Psiarb}), where $ \ket{\varphi_n}= \ket{\varphi}^{\otimes n}$. In this section, we will examine how the encryption that samples unitaries with respect to Haar measure on the unitary group $U(2)$ works. Let us for now ignore the fact that such scenario is impractical for encryption of quantum information, because it assumes existence of (uncountably) infinite classical key shared between Alice and Bob, however, the goal is to show that such encryption constitutes a valid private quantum channel also in the case of the considered multi-photon source.

From the point of view of the eavesdropper the encryption channel
$\mathcal{E}_{\rm Haar}$ maps the overall density matrix $\rho=\isom(\ket{\Psi}\bra{\Psi})$ into
\begin{align}
\nonumber
\rho'&=\mathcal{E}_{\rm Haar}(\rho)
=\int_{U(2)} dU L(U)\rho L(U)^\dagger \\
\label{eq:haar_encryption}
&=\int_{U(2)}dU  \sum_{m,n=0}^\infty c_m c^*_n U^{\otimes m}\ket{\varphi^{\otimes m}}\bra{\varphi^{\otimes n}} (U^\dagger)^{\otimes n}\,.
\end{align}
By construction $\rho'$ commutes with any operator from the representation $L(U)$ and Schur's lemma implies that $\rho'$ is block diagonal in the irreducible subspaces. In our case these are exactly subspaces $\hilb{K}^+_{n}$ of $\hilb{K}$, thus,
\begin{align}
\label{eq:eaction}
\rho'= \sum_{n=0}^\infty  {\rm tr}[\rho\Pi_n] \frac{1}{n+1}\Pi_n
= \sum_{n=0}^\infty |c_n|^2 \frac{1}{n+1}\Pi_n\, ,
\end{align}
where $\Pi_n$ are the projections onto subspaces $\hilb{K}^+_{n}$. Although, the state $\rho'$ carries some information about the imperfections of the source, contained in the coefficients $c_n$, it does not posses any information about the polarization degrees of freedom. Consequently, the encryption is secure \tcb{(in accordance with \cite{2000ambainis})}, thus the Haar measure defines a valid quantum private channel.

In practice, the energy of a source is limited, so it is reasonable to put some upper bound $N$ on the number of photons that the source might generate. In such setting it is reasonable to require that the encryption channel acts in the same way as channel $\mathcal{E}_{Haar}$ on the subspace $\hilb{K}(N)\equiv \oplus_{n=0}^N \hilb{K}^+_{n}$ corresponding to at most $N$ photons. Since the channel $\mathcal{E}_{Haar}$ does not mix subspaces $\hilb{K}^+_{n}$ its restriction $\mathcal{E}_{N,{\rm Haar}}$ to operators on $\hilb{K}(N)$ is well defined.

Due to Choi-Jamiolkowski isomorphism \cite{choi} a channel $\mathcal{E}_{N,{\rm Haar}}:\hilb{L}(\hilb{K}(N))\rightarrow \hilb{L}(\hilb{K}(N))$ can be represented as an operator $H_N$
\begin{align}
\label{eq:choidef}
H_N=\mathcal{E}_{N,{\rm Haar}}\otimes \mathcal{I} (\ket{\omega}\bra{\omega}),
\end{align}
where $\ket{\omega}=\sum_{i=1}^{\dim \hilb{K}(N)} \ket{i}\otimes\ket{i}\in \hilb{K}(N)\otimes\hilb{K}(N)$ is an unnormalized maximally entangled state. Without loss of generality we will assume that the basis $\{\ket{i}\}_{i=1}^{\dim \hilb{K}(N)}$ respects the subspaces $\hilb{K}^+_{n}$. In particular, it is a union of orthonormal bases $\{\ket{e^n_k}\}_{k=0}^n$ (for $n=0,\dots ,N$) associated with subspaces $\hilb{K}^+_{n}$. Define $\ket{\omega_n}=\sum_{k=0}^n \ket{e^n_k}\otimes\ket{e^n_k}\in \hilb{K}(N)\otimes\hilb{K}(N)$. Then
\begin{align}
\label{eq:omegaexp}
\ket{\omega}=\sum_{n=0}^{N} \ket{\omega_n} \,.
\end{align}
Using Eqs. (\ref{eq:eaction}), (\ref{eq:choidef}) we obtain
\begin{align}
H_N &= \sum_{m,n,k=0}^{N}   \frac{1}{k+1}\Pi_k \otimes  {\rm tr}_1[\ket{\omega_m}\bra{\omega_n} (\Pi_k\otimes I)] \\
&= \sum_{n=0}^{N}   \frac{1}{n+1}  \Pi_n\otimes \Pi_n\,,
\end{align}
where ${\rm tr}_1$ denotes the partial trace over the first part of the tensor product and we used $\Pi_k\otimes I\ket{\omega_l}=\delta_{kl}\ket{\omega_l}$ and ${\rm tr}_1[\ket{\omega_n}\bra{\omega_n}]=\Pi_n$. In other words,
$\forall m,n=0,\ldots,N$
\begin{align}
\label{eq:choipodmienka}
\mathcal{E}_{N,{\rm Haar}}\otimes\mathcal{I}(\ket{\omega_m}\bra{\omega_n})
=\frac{\delta_{mn}}{n+1} \Pi_n\otimes \Pi_n \,.
\end{align}

In conclusion, in this section we have shown that  i) unitary changes of polarisation chosen randomly according to Haar measure form a private quantum channel in multi-photon settings ii) suitable encryption channel should have the same Choi operator as $\mathcal{E}_{N,{\rm Haar}}$ if at most $N$ photons are expected to be simultaneously generated by the source.

\section{Unitary k-designs and multiphoton sources}
We have shown that Haar measure sampling of polarization transformations enable us to design a private quantum channel for multiphoton sources. However, such encryption is very ``impractical", because it assumes existence of continous key between the sender and the receiver. In this section we will address the question whether $k$-designs can be used to implement private quantum channels for multiphoton sources. This question is motivatied by the following observation.

It is known that channels generated by $k$-fold tensor products of Haar distributed unitary channels ($U^{\otimes k}$) are not mapping the state space of $k$-partite system into a single point, thus, 
the related k-designs are not sufficient to encrypt arbitrary k-partite state. However, when restricted to completely symmetric states $k$-designs define a valid private quantum channel. The indistinguishability of photons allows us to represent the multiphoton states by completely symmetric states, thus, $k$-designs might be relevant for their encryption.

It is the goal of a unitary $k$-design (for review see \cite{2011zauner})
to mimic the average effect of Haar distributed unitary channels. In particular, a set $\{U_j\}_{j=1}^M$ is a \emph{qubit unitary $k$--design} $\mathcal{F}_k$ if for all states $\xi$ of $k$ qubits (associated with a Hilbert space $\hilb{Q}_k$)
\begin{align}
\label{eq:kdesigndef}
\mathcal{F}_k(\xi)\equiv\frac{1}{|M|}\sum_{j=1}^M U_j^{\otimes k}\xi U_j^{\dagger\otimes k}=\int_U U^{\otimes k}\xi U^{\dagger\otimes k}dU\,.
\end{align}
Let us stress that any $k$-design is also $(k-1)$--design, i.e.
\begin{align}
  \nonumber
  \mathcal{F}_{k-1}(\xi)
 & \equiv\frac{1}{|M|}\sum_{j=1}^M U_j^{\otimes (k-1)}\xi U_j^{\dagger\otimes (k-1)}\\
  \nonumber
 & =\int_U U^{\otimes (k-1)}\xi U^{\dagger\otimes (k-1)}dU\,,
\end{align}
for all $\xi\in\hilb{L}(\hilb{Q}_{k-1})$.

For the multiphoton case we are asking whether sampling of the qubit unitary k-designs resembles the sampling over the whole Haar measure when the energy of the multiphoton source is restricted, i.e. we want to see whether (Eq.\eqref{eq:haar_encryption}) for all $\varrho\in{\cal S}({\cal K})(k)$
\begin{align}
\sum_j q_j L(U_j)\varrho L(U_j)^\dagger={\cal E}_{k,Haar}(\varrho)
\end{align}
for some $k$-design $\{q_j,U_j\}$.

In what follows we will analyze the channel $\mathcal{F}_k$ induced by qubit unitary
$k$--design in a basis 
that is induced in $\hilb{Q}_k$ by representation $U \mapsto U^{\otimes k}$ of group $SU(2)$. 
Let us denote by $\hilb{M}_s$ the $2s+1$-dimensional Hilbert space of spin $s$ irreducible representation of $SU(2)$, where $s=0,\frac{1}{2},1,\frac{3}{2},2,\frac{5}{2},\dots$. Set either $s_0=0$, or $s_0=1/2$ if $k$ is even, or odd, respectively.
The whole Hilbert space $\hilb{Q}_k$ can be 
decomposed as 
\begin{align}
\label{eq:defmk}
\hilb{Q}_k=\bigoplus_{s=s_0}^{k/2} \hilb{M}_s \otimes \mathbb{C}^{m_s},
\end{align}
where $\mathbb{C}^{m_s}$ is the multiplicity space of the dimension
$m_s=\frac{2s+1}{k/2+s+1} \binom{k}{k/2+s}$ (see \cite{1999cirac}).
In such basis $k$-fold tensor product of an arbitrary qubit operator $U$ is expressed as 
\begin{align}
\label{eq:su2decomp}
U^{\otimes k}=\bigoplus_{s=s_0}^{k/2} U_s \otimes I_{m_s}.
\end{align}
where $U_s\in \hilb{L}(\hilb{M}_s)$ and $I_{m_s}$ denotes the unit operator
on $\mathbb{C}^{m_s}$. Especially, for $U=U_j$ we will use the notation
$U_j^{\otimes k}=\bigoplus_{s=s_0}^{k/2} U^{(j)}_{s} \otimes I_{m_s}$. We see
that both $k$--design and the Haar distributed unitaries act trivially
in the multiplicity spaces $\mathbb{C}^{m_s}$.

Thanks to this fact there exist infinitely many subspaces $\mathcal{W}$ of $\hilb{Q}_k$, which are isomorphic to $\bigoplus_{s=s_0}^{k/2} \hilb{K}^+_{2s}$ and where the representation of the group $U(2)$ acts in the same way. More precisely, we choose a one dimensional subspace $\pi_{s}$ in every space $\mathbb{C}^{m_s}$, thus we have $\mathcal{W}=\bigoplus_{s=s_0}^{k/2} \hilb{M}_s \otimes \pi_s$. This naturally provides the described isomorphism if basis of the irreducible subspaces $\hilb{M}_s$ and $\hilb{K}^+_{2s}$ are suitably paired, i.e. $\ket{e^s_j}\in \hilb{K}^+_{2s}$ 
corresponds to $\ket{f^s_j}\in\hilb{M}_s$ for suitably chosen orthonormal basis $\{\ket{f^s_j}\}_{j=1}^{2s+1}$.
In order to keep the notation simple we will not distinguish $\hilb{M}_s \otimes \pi_s$
and $\hilb{M}_s$, because they are isomorphic. It follows from Eq.~(\ref{eq:su2decomp})
that $U^{\otimes k}$ acts in $\mathcal{W}$ as a unitary transformation $\bigoplus_{s=s_0}^{k/2} U_s$. Since $k$-design $\mathcal{F}_k$ is a mixture of unitary channels $U_j^{\otimes k}$ 
we can conclude that it induces a quantum channel 
$\mathcal{F}_{k,\mathcal{W}} (\,\xi\,)=\sum_j p_j (\bigoplus_{s=s_0}^{k/2} U_s^{(j)})\,\xi\, (\bigoplus_{s'=s_0}^{k/2} U_{s'}^{(j)\dagger})$
acting on the subspace $\mathcal{W}\subset\hilb{Q}_k$.

Using the above mentioned isomorphism between $\hilb{K}^{+}_{2r}$ and $\hilb{M}_r$
it follows that (see Eq.\eqref{eq:omegaexp})
\begin{align}
\ket{\omega_r}=\sum_{j=1}^{2r+1} \ket{f^r_j}\otimes \ket{f^r_j}
\end{align}
is the unnormalized maximally entangled state on
$\mathcal{M}_r\otimes\mathcal{M}_r$, thus,
$\ket{\Omega}=\bigoplus_{r=s_0}^{k/2} \ket{\omega_r}=\sum_{r=s_0}^{k/2}\ket{\omega_r}$
is the unnormalized maximally entangled state on $\hilb{W}\otimes\hilb{W}$.
Consequently, the corresponding Choi-Jamiolkowski state of $\mathcal{F}_{k,\mathcal{W}}$ reads
\begin{align}
  \nonumber
 F_{k,\mathcal{W}}&=(\mathcal{F}_{k,\mathcal{W}}\otimes \mathcal{I})\ket{\Omega}\bra{\Omega}=(\mathcal{F}_{k,\mathcal{W}}\otimes \mathcal{I})\left(\sum_{r,t=s_0}^{k/2} \ket{\omega_r}\bra{\omega_t}\right)\\
 \nonumber
&= \int_{U(2)}dU \sum_{s} (U_s \otimes I_{2s+1}) \ket{\omega_s}\bra{\omega_s} (U_s \otimes I_{2s+1})^\dagger\\
&= \sum_{s=s_0}^{k/2} \frac{1}{2s+1} I_{2s+1} \otimes I_{2s+1} \,,
\end{align}
where we using the same Schur's lemma argument as in section \ref{haarandsource} conclude that the Haar averaging erases the parts of the input state, which map between subspaces of different irreducible representations (irrep) while for the same irreps it creates a multiple of the completely mixed state $\frac{1}{2s+1}I_{2s+1}\in \hilb{L}(\hilb{M}_{s})$.

\begin{figure}[t]
\centerline{\includegraphics[height=3.6cm,width=7cm]{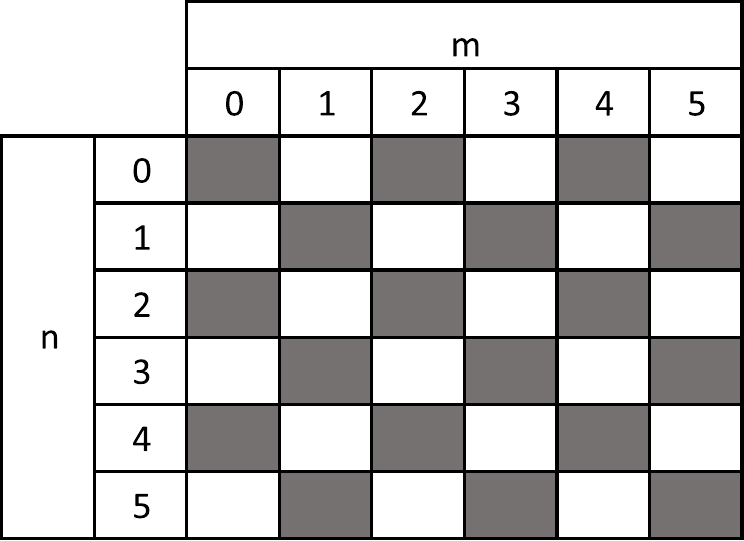}}
\caption{The table schematically illustrates 
for which blocks (depicted in gray) the Choi-Jamiolkowski operators of $\mathcal{E}_{N,{\rm Haar}}$ and $\mathcal{F}_{k,\mathcal{W}}$ coincide in case of 
a qubit $5$--design.}
\label{fig:table1}
\end{figure}

It is useful to define subspaces $\hilb{K}_{\rm even}(n)=\bigoplus_{s=0}^{n/2} \hilb{K}^+_{2s}$ and
$\hilb{K}_{\rm odd}(n)=\bigoplus_{s=0}^{(n-1)/2} \hilb{K}^+_{2s+1}$, because 
the subspace $\hilb{W}$ is either isomorphic to the subspace $\hilb{K}_{\rm even}(k)$ (even number of photons) or to $\hilb{K}_{\rm odd}(k)$ (odd number of photons). 
Suppose $\hilb{W}=\hilb{K}_{\rm even}(k)$ (meaning that $k$ is even and $s_0=0$). Then
clearly $F_{k,\mathcal{W}}=H_k$ on the subspace
$\mathcal{W}\simeq\bigoplus_{s=0}^{k/2} \hilb{K}^+_{2s}\equiv\hilb{K}_{\rm even}(k)
\subset \hilb{K}(k)$. Moreover, for $(k-1)$--design (induced by the considered
$k$--design) we find (following all the previous steps) that
$F_{k-1}=H_k$ on $\hilb{K}_{\rm odd}(k-1)\subset\hilb{K}(k)$.

In conclusion, if elements of qubit $k$--design (whether $k$ is even, or odd) determine polarization rotations of 
multi-photon source (generating at most $k$ photons) this constitutes
a private quantum channel for subspaces of odd and even number of photons (see Fig.~(\ref{fig:table1}). 
In particular, we have shown that for all qubit states $\ket{\varphi},\ket{\psi}$
the following identity holds  
\begin{align}
\label{eq:conemame2}
\mathcal{F}_k(\ket{\varphi^{\otimes m}}\bra{\psi^{\otimes n}})=\frac{1}{|M|}\sum_{j=1}^M U_j^{\otimes m}  \ket{\varphi^{\otimes m}}\bra{\psi^{\otimes n}} U_j^{\dagger\otimes n}=0 \,,
\end{align}
if $m$ and $n$ have the same parity (both even, or both odd).
However, for states of undefined parity 
it is not clear if $k$--design would erase coherences
between odd and even photon number subspaces of $\hilb{K}(k)$. In what follows we will show that this is not the case.

Consider the 12-element Clifford 3-design $\mathcal{E}_{\rm Clifford}$ \cite{Webb16}. Direct calculation shows (see Appendix \ref{sec:app1}) that off-diagonal block of its Choi-Jamiolkowski state 
$(\mathcal{E}_{\rm Clifford}\otimes \mathcal{I})\ket{\omega_{\tcb{2}}}\bra{\omega_{1}}$
is nonzero, thus  the state 
$(\mathcal{E}_{\rm Clifford}(\ket{\Psi}\bra{\Psi})$
still depends on the polarisation parameters $\alpha$ and $\beta$. 
In conclusion, $k$-designs do not necessarily constitute private quantum channels.

\section{Discussion}
In this paper we have shown that Haar sampled unitaries constitute  private quantum channel that remains secure even when photon source generates more copies of identically prepared polarized photons (multi-photon source). However, unlike in the case of the single-photon sources encryption, a general $k$--design does not have this property.

Let us now discuss conditions on a multi-photon source under which its
secure encryption by polarization rotations forming a $N$--design is possible. Clearly this happens when a source is ``parity stable", i.e. it is
producing states with nonzero amplitudes in only even (or only odd)
photon number subspaces. An example could be a Type I
spontaneous parametric down conversion source, which produces only even photon
number terms of the same polarization. For a general multi-photon source
we may think of its combination with selective erasure channel eliminating
coherences between subspaces of different parity while keeping the other subspaces undisturbed. An example is the non-destructive photon number measurement. In practice, it can be realized by using spontaneous parametric down-conversion source, where in contrast to previous case we would use signal and idler photons separately. Measuring the number of photons (or just parity) of the idler photon, the signal photon would be collapsed to a state with zero coherences between different photon number subspaces.

One can also argue that if the encryption is restricted to be realized by linear optics polarization changes (i.e. by randomly choosing unitary transformations $L(U)$) then the only encryption channel secure without additional restrictions is given by $\mathcal{E}_{\rm Haar}$. The idea is that no matter with which polarization we start, the output state must be the same. Since the operations $L(U)$ can only rotate the polarization the only possibility how the random mixture of rotations gives always the same output is when they always yield a state covariant with respect to rotations. Since linear optical elements do not change the photon number we get the uniqueness.

\begin{acknowledgments}
This project is supported by project MUNI/G/1211/2017 (GRUPIK). M.S. was supported by The Ministry of Education, Youth and Sports of the Czech Republic from the National Programme of Sustainability (NPU II); project IT4Innovations excellence in science - LQ1602. M.Z. acknowledges the support by the QuantERA project HiPhoP (Project ID No. 731473), project APVV-18-0518 (OPTIQUTE) and VEGA 2/0173/17 (MAXAP).
\end{acknowledgments}

\appendix
\section{Clifford $3$-design}
\label{sec:app1}
The goal of this section is to demonstrate existence of a $k$-design, which if used to 
determine polarization rotations of a multi-photon source does not create a private quantum channel.
Consider a channel ${\cal E}_{\rm Clifford}(\varrho)=\frac{1}{12}\sum_j L(U_j)\varrho L(U_j)^\dagger$ 
generated by the 12-element Clifford 3-design \cite{Gross07,Webb16}. \tcb{The $3$-design $\{U_j\}_{j=1}^{12}$ is formed} by identity, three Pauli operators ($\sigma_0=I, \sigma_x,\sigma_y,\sigma_z$) and eight unitary operators 
\begin{align}
\label{eq:tridizajnmatice2}
\nonumber
U_{klm}=&\exp [\imath \frac{2\pi}{3} \vec{n}_{klm}\cdot \vec{\sigma}] 
\end{align}
$\vec{n}_{klm}=\frac{1}{\sqrt{3}}\left((-1)^k,(-1)^l,(-1)^m\right)$ and $k,l,m=0,1$.
Vectors $\vec{n}_{klm}$ form vertices of a cube.
\tcb{
The above mentioned $12$ operators in arbitrary order will be further addressed as $U_j$.
Although above we described a $3$-design we will only use it as a qubit $2$-design and $1$-design. 
If ${\cal E}_{\rm Clifford}$ was a private quantum channel then its Choi operator would be the same as for ${\cal E}_{N, Haar}$ (see Eq. (\ref{eq:choipodmienka})). In particular, ${\cal E}_{N, Haar}(\ket{\omega_2}\bra{\omega_1})=0$ and we will show that ${\cal E}_{N, \rm Clifford}(\ket{\omega_2}\bra{\omega_1})\neq 0$, i.e. the block of the Choi operator that describes transformation between $2$ photon and $1$ photon subspaces of the Fock space (or equivalently using the isomorphism between 2/2 and 1/2 spin irreps) is non zero. 
For the output space it is more practical to calculate in the space of two qubits, where spin $2/2$ irrep is naturally embedded in the symmetric subspace $P_{sym}$ and elements of the design $U_j$ are represented by $U^{\otimes 2}_j$. Thus, we will use operator 
$T=\ket{2,0}\bra{00}+\ket{1,1}\frac{1}{\sqrt{2}}(\bra{01}+\bra{10})+\ket{0,2}\bra{11}$ to transport this three dimensional subspace of $2$-qubits back into three dimensional $2$-photon subspace of the Fock space. 
Similarly, a single qubit $2$-dimensional Hilbert space is transferred by operator $S=\ket{1,0}\bra{0}+\ket{0,1}\bra{1}$.} 
Then, evaluating the block  
\begin{align}
\nonumber
C_{\tcb{\frac{2}{2}}\frac12}&\equiv (\mathcal{E}_{\rm Clifford}\otimes \mathcal{I})\ket{\omega_{\tcb{2}}}\bra{\omega_{1}} \\
&=\frac{1}{12}\sum_j (U_{\tcb{2/2}}^{(j)}\otimes I_{\tcb{2/2}}) \ket{\omega_{\tcb{2}}}\bra{\omega_{1}}  (U_{1/2}^{(j)\dagger}\otimes I_{1/2}) \nonumber\\
&=\frac{1}{12}\sum_j (T \!\otimes \! T)(U_j^{\otimes 2}\!\otimes \!I) \ket{\omega_+}\bra{\phi_+}  (U_j^{\dagger}\!\otimes \!I) \tcb{(S^\dagger \!\otimes \!S^\dagger)} \nonumber
\end{align}
results in
\begin{align}
C_{\tcb{\frac{2}{2}}\frac12}=
\left(
\begin{array}{cccc}
a & 0 & 0 & -b \\
0 & c & 0 & 0 \\
0 & b & -b^* & 0 \\
0 & 0 & c & 0 \\
b^* & -b^* & b & b \\
0 & c & 0 & 0 \\
0 & b & -b^* & 0 \\
0 & 0 & c & 0 \\
-b^* & 0 & 0 & a^* 
\end{array} 
\right) \neq 0
\end{align}
where $a=\frac{3+\imath}{12}$, $b=\frac{1+\imath}{12}$, $c=\frac{1}{3 \sqrt{2}}$ and we used
$\ket{\phi_+}=\frac{1}{\sqrt{2}}(\ket{00}+\ket{11})$, 
$\ket{\omega_+}=\ket{00}^{\otimes 2} + \frac{1}{2}(\ket{01}+\ket{10})^{\otimes 2}+ \ket{11}^{\otimes 2}$.
\tcb{
As the input vector $\ket{\omega_+}$ belongs to $P_{sym}\otimes P_{sym}$ subspace it is clear that a four qubit unitary (16x16 matrix) $U^{\otimes 2}_j\otimes I$ will leave $\ket{\omega_+}$ in that subspace and after operator $ T\otimes T$ we get something from a $9=3\times 3$ dimensional subspace corresponding to tensor product of two $2$-photon Fock subspaces. On the input the operator $C_{\frac{2}{2}\frac12}$ is nonzero only for $4=2\times 2$ dimensional subspace formed by tensor product of two 1-photon subspaces.
Thus, in corresponding basis operator $C_{\frac{2}{2}\frac12}$ has $9\times 4$ matrix, which is evidently nonzero and so we showed that ${\cal E}_{Clifford}(\ket{\omega_2}\bra{\omega_1})\neq 0$. We conclude that ${\cal E}_{\rm Clifford}$ does not constitute a private quantum channel.
}

\section{Fixed parity example}
\tcb{
In this appendix we will discuss an explicit example illustrating that  
for pure polarization states of a fixed parity the encoding provided by Clifford design
is secure, thus, no information on the polarisation can be extracted from the resulting ciphertext. 
}

\tcb{Suppose the multi-photon source has produced a state (plaintext) 
\begin{align}
\label{eq:psi02}
\ket{\psi}&=c\ket{0,0}  + \sqrt{1-|c|^2}(\alpha a^\dagger_x + \beta a^\dagger_y)^2\ket{0,0} \nonumber \\
        &=c\ket{0,0}+ \sqrt{1-|c|^2} (\alpha^2 \ket{2,0} + \sqrt{2}\alpha\beta\ket{1,1} +\beta^2 \ket{0,2} \nonumber \\
        &=c\ket{\psi_0}+ \sqrt{1-|c|^2} \ket{\psi_2}
\end{align}
which is a superposition of vacuum and 2-photons in a fixed polarization given by amplitudes $\alpha, \beta$. After the encryption by ${\cal E}_{\rm Clifford}$ we obtain:
\begin{align}
{\cal E}_{\rm Clifford}(\ket{\psi}\bra{\psi})=&|c|^2 \ket{0,0}\bra{0,0} \nonumber \\
&+ c\sqrt{1-|c|^2} {\cal E}_{\rm Clifford}(\ket{\psi_0}\bra{\psi_2}) \nonumber \\
&+ c^* \sqrt{1-|c|^2} {\cal E}_{\rm Clifford}(\ket{\psi_2}\bra{\psi_0}) \nonumber \\
&+  (1-|c|^2) {\cal E}_{\rm Clifford}(\ket{\psi_2}\bra{\psi_2}),\nonumber 
\end{align}
where we used that the vacuum state is unaffected by polarization rotations.
For the second and third term we need to calculate
\begin{align}
\frac{1}{12}\sum_{j=1}^{12} U^{(j)}_{2/2} \ket{\psi_2}&=T\;( \frac{1}{12}\sum_{j=1}^{12}  U_j^{\otimes 2}) \ket{\varphi_2} \nonumber \\
        &=0,
\end{align}
where $\ket{\varphi_2}=\alpha^2 \ket{00} + \alpha\beta(\ket{01}+\ket{10}) +\beta^2 \ket{11}$ and $T$ is defined Appendix \ref{sec:app1}. By direct evaluation of the sum we find that it equals to the projector onto 2-qubit antisymmetric subspace and this nulifies the symmetric state $\ket{\varphi_2}$ on the right.
We evaluate the last term in similar fashion.
\begin{align}
{\cal E}_{\rm Clifford}(\ket{\psi_2}\bra{\psi_2})&=\frac{1}{12}\sum_{j=1}^{12} U^{(j)}_{2/2} \ket{\psi_2}\bra{\psi_2}(U^{(j)}_{2/2})^\dagger \nonumber \\
&=T\; \frac{1}{12}\sum_{j=1}^12  U_j^{\otimes 2} \ket{\varphi_2}\bra{\varphi_2} (U_j^{\otimes 2})^\dagger \; T^\dagger \nonumber \\
        &=\frac{1}{3} \Pi_2,
\end{align}
We used that qubit 2-design is mapping symmetric pure states to a third of a projector onto the symmetric subspace, which is by $T$ mapped into projector onto $2$-photon subspace of the Fock space.
Putting all together we obtained:
\begin{align}
{\cal E}_{\rm Clifford}(\ket{\psi}\bra{\psi})=&|c|^2 \ket{0,0}\bra{0,0} \nonumber \\
&+  (1-|c|^2)\frac{1}{3} \Pi_2,\nonumber 
\end{align}
As expected we see that information on polarization amplitudes is completely erased, only the imperfections of the source represented here by coefficient $c$ remain.
}

\section{Note on security definition of private quantum channels}
\label{sec:explanation_of_definition}


\tcb{
Unconditionally secure (in fact information-theoretically secure) classical encryption was formally introduced in the seminal paper by Claude Shannon in \cite{shannon1949}, where he also shows that one-time pad encryption system fulfills this definition.
}

\tcb{
In the traditional terminology of cryptography, the sender chooses a message (called plaintext) she wants to send. The set of all possible plaintexts (e.g. the set of all bit strings of length 64) is publicly known and is a part of the specification of an encryption system. In order to transfer the message securely, sender and receiver must have some advantage over the eavesdropper, usually in the form of a pre-shared secret bit string called key. The set of all possible keys (once again e.g. the set of all bit strings of length 64) is public and is a part of the specification of the encryption system. Sender and receiver choose the key randomly from this set, according to a public and pre-agreed probability distribution (most of the time uniform).
}

\tcb{
Sender then transforms the plaintext, using an encryption function, into a ciphertext. Encryption function is publicly known binary function, taking plaintext and key as inputs. The encryption function is public and part of the specification of the encryption system.
Ciphertext is then transmitted via an insecure channel to the receiver, with the expectation that it will be observed by an eavesdropper. When the receiver obtains the message, he will use a decryption function and key to obtain the plaintext. Once again, decryption function is public and is a part of the specification of the encryption system.
}

\tcb{
The definition of secrecy, when stated informally, requires that observing the ciphertext gives no extra information about the plaintext to the eavesdropper. By ``extra information" we mean that the eavesdropper may have some a priori information regarding plaintext, e.g. that it is a meaningful text in English language encoded into a bit string using ASCII character encoding.
}


\tcb{
Let us denote the set of all plaintexts by $P$, set of all keys by $K$ together with some probability distribution on $K$ giving a rise to the random variable ${\bf K}$, and a set of all ciphertexts by $C$. Plaintexts and ciphertexts are mutually bound via the encryption and decryption function by $c=e(p,k)$ and $p=d(c,k)$. These two functions together with probability distribution on $K$ give rise to conditional distributions (random variables) ${\rm prob}(P=p|C=c)$ and ${\rm prob}(C=c|P=p)$. Formally, the encryption systems is the five--tuple $(P,C,{\bf K},e,d)$. The standard form of the perfect secrecy condition then reads
$$
\exists x\ s.t.\ \forall c\in C, \forall p\in P\ {\rm prob}(P=p|C=c)=x
$$
The crucial property for the secrecy is that regardless of what ciphertext ``$c$" is being transmitted, the probability distribution ${\rm prob}(P=p|C=c)$ on plaintexts induced by any ciphertext ``$c$" is the same for all ciphertexts. Regardless of what ciphertext the eavesdropper observes, his estimate of the plaintext remains the same.
}


\tcb{
In order to understand the definition of Private quantum channel, we use
an equivalent statement
$$
\exists y\ s.t.\ \forall c\in C, \forall p\in P\ {\rm prob}(C=c|P=p)=y.
$$
This can be further transformed to
$$
\exists q_0\ s.t.\ \forall p\in P\ {\rm prob}(C=c|P=p)=q_0(c),
$$
with $q_o$ being a (conditional) probability distribution on ciphertexts. This formulation accents the fact that the conditional probability distribution ${\rm prob}(C=c|P=p)$ describing probability distribution of $C$ conditioned by a particular $p$ is fixed. It remains to note that in the definition of private quantum channel, the average state $\rho^{(0)}$ is the equivalent of conditional probability distribution $q_0$ on the ciphertexts and the set of all states plays the role of the set of plaintexts. This gives us exactly the definition 
of the PQC:
\begin{equation*}
    \exists \rho^{(0)}\ s.t.\ \forall\rho\ \sum_i p_i U_i\rho U_i^\dagger=\rho^{(0)},
\end{equation*}
where $\rho$ is the ciphertext being transmitted, and $\sum_i p_i U_i\rho U_i^\dagger$ is our prediction of the ciphertext given the particular plaintext $\rho$. The PQC definition says this prediction is independent of the plaintext in accordance with the classical definition of security. If this condition is not satisfied, then there are at least two different plaintexts giving rise to different average states that can be in principle discriminated.
}

\end{document}